\documentclass[twocolumn]{revtex4}
\usepackage{amsmath}

\begin{document}

\title{Inversion formula and Parsval theorem for complex continuous wavelet
transforms studied by entangled state representation{\small \thanks{%
Project supported by the National Natural Science Foundation of China (Grant
Nos 10475056 and 10775097).}}}
\author{{\small Li-yun Hu}$^{\ast 1,2}${\small \ and Hong-yi Fan}$^{2}$}
\affiliation{$^{1}${\small College of Physics and Communication Electronics, Jiangxi
Normal University, Nanchang 330022, China}\\
$^{2}${\small Department of Physics, Shanghai Jiao Tong University,
Shanghai, 200030, China}\\
$^{\ast }${\small Corresponding author. E-mail: hlyun2008@126.com (L-Y Hu)}}

\begin{abstract}
In a preceding Letter (Opt. Lett. 32, 554 (2007)) we have proposed complex
continuous wavelet transforms (CCWTs) and found Laguerre--Gaussian mother
wavelets family. In this work we present the inversion formula and Parsval
theorem for CCWT by virtue of the entangled state representation, which
makes the CCWT theory complete. A new orthogonal property of mother wavelet
in parameter space is revealed.

\end{abstract}

\maketitle

\section{Introduction}

Wavelet transforms (WTs) are very useful in signal analysis and detection
\cite{1,2,3} since it can overcome the shortcomings of nonlocality behavior
of classical Fourier analysis and thus enriches the theory of Fourier optics
\cite{4}. The continuous WT of a signal function $f\left( x\right) \in
L^{2}\left( R\right) $ by a mother wavelet $\psi \left( x\right) $
(restricted by the admissibility condition $\int_{-\infty }^{\infty }\psi
\left( x\right) dx=0$) is defined by%
\begin{equation}
W_{\psi }f\left( \mu ,s\right) =\frac{1}{\sqrt{\mu }}\int_{-\infty }^{\infty
}f\left( x\right) \psi ^{\ast }\left( \frac{x-s}{\mu }\right) dx,
\label{1.1}
\end{equation}%
where $\mu $ $\left( >0\right) $ is a scaling parameter and $s\left( \in
R\right) $\ is a translation parameter. The inversion of (\ref{1.1}) is%
\begin{equation}
f\left( x\right) =\frac{1}{C_{\psi }}\int_{0}^{\infty }\frac{d\mu }{\mu ^{2}}%
\int_{-\infty }^{\infty }W_{\psi }f\left( \mu ,s\right) \psi \left( \frac{x-s%
}{\mu }\right) \frac{ds}{\sqrt{\mu }},  \label{1.2}
\end{equation}%
where $C_{\psi }=\int_{0}^{\infty }\frac{\left\vert \psi \left( p\right)
\right\vert ^{2}}{p}dp<\infty $ and $\psi \left( p\right) $ is the Fourier
transform of $\psi \left( x\right) ,$ for proving (\ref{1.2}) we have
employed the Dirac's representation theory \cite{5}, which has the merit of
rigour and simplicity.

In Ref.\cite{6,7}, Fan and Lu have linked the one-dimensional (1D) WT with
the unitary transform (squeezing and dispacement) in quantum mechanics,
i.e., expressing the WT as a matrix element of the single-mode
squeezing-displacing operator between the mother wavelet state vector $%
\left\langle \psi \right\vert $ and the state vector to be transformed, such
that the admissibility condition for mother wavelets is examined in the
context of quantum mechanics, in so doing a family of the Hermite--Gaussian
mother waveletes are found. Further, by introducing the bipartite entangled
state representation $\left\vert \eta \right\rangle $ \cite{8}
\begin{equation}
\left\vert \eta \right\rangle =\exp \left( -\frac{1}{2}\left\vert \eta
\right\vert ^{2}+\eta a_{1}^{\dagger }-\eta ^{\ast }a_{2}^{\dagger
}+a_{1}^{\dagger }a_{2}^{\dagger }\right) \left\vert 00\right\rangle ,
\label{1.4}
\end{equation}%
Fan and Lu then proposed the continuous complex wavelet transforms (CCWT)
for $g\left( \eta \right) \equiv \left\langle \eta \right\vert \left.
g\right\rangle ,\ $%
\begin{equation}
W_{\psi }g\left( \mu ,\kappa \right) =\frac{1}{\mu }\int \frac{d^{2}\eta }{%
\pi }g\left( \eta \right) \psi ^{\ast }\left( \frac{\eta -\kappa }{\mu }%
\right) ,  \label{1.5}
\end{equation}%
where $\kappa \in C.$ Correspondingly, the admissibility condition for
mother wavelets, $\int \frac{d^{2}\eta }{2\pi }\psi \left( \eta \right) =0,$
is examined in the entangled state representations and a family of new
mother wavelets (named the \textbf{Laguerre--Gaussian} wavelets) are found
to match the CCWT \cite{9}, i.e., the qualified mother wavelets $\psi \left(
\eta \right) $ satisfying the admissibility condition can be expressed as
the function of $\left\vert \eta \right\vert ,$%
\begin{eqnarray}
\psi \left( \eta \right) &=&e^{-\left\vert \eta \right\vert
^{2}/2}\sum_{n=0}^{\infty }K_{n,n}(-1)^{n}H_{n,n}^{\ast }\left( \eta ,\eta
^{\ast }\right)  \notag \\
&=&e^{-\left\vert \eta \right\vert ^{2}/2}\sum_{n=0}^{\infty
}n!K_{n,n}L_{n}\left( \left\vert \eta \right\vert ^{2}\right) ,  \label{23}
\end{eqnarray}%
where $L_{n}\left( x\right) $ is the\ Laguerre\ function and $H_{m,n}\left(
x,y\right) $ is the two-variable Hermite polynomial \cite{10}, whose
generating function is
\begin{equation}
H_{m,n}\left( x,y\right) =\frac{\partial ^{m+n}}{\partial t^{\prime
n}\partial t^{m}}\exp \left[ -tt^{\prime }+tx+t^{\prime }y\right]
_{t=t^{\prime }=0}.  \label{24}
\end{equation}%
We emphasize that the CCWT differs from the direct product of two 1D WTs
since the squeezing transform involved in (\ref{1.5}) is in two-mode%
\begin{equation}
\frac{1}{\mu }\psi ^{\ast }\left( \frac{\eta -\kappa }{\mu }\right) =\frac{1%
}{\mu }\left\langle \psi \right. \left\vert \frac{\eta -\kappa }{\mu }%
\right\rangle =\left\langle \psi \right\vert S_{2}\left\vert \eta -\kappa
\right\rangle ,  \label{1.7}
\end{equation}%
where $S_{2}$ is the two-mode squeezing operator $S_{2}=\exp
[(a_{1}^{\dagger }a_{2}^{\dagger }-a_{1}a_{2})\ln \mu ]$ \cite{11,12}, which
is in sharp contrast to the direct-product of two single-mode squeezing
(dilation) operators, and the two-mode squeezed state is simultaneously an
entangled state.

In order to complete the CCWT theory, we must ask if the corresponding
Parseval theorem exists. This is important since the inversion formula of
CCWT may appear as a lemma of this theorem. We shall solve this issue by
virtue of the merits of entangled state in quantum mechanics, to be more
specific, we shall use the property that the two-mode squeezing operator has
its natural representation in the entangled state basis (see (\ref{1.13})
below). Noting that CCWT involves two-mode squeezing transform, so the
corresponding Parseval theorem differs from that of the direct-product of
two 1D wavelet transforms, too.

\section{The quantum mechanical version of CCWT}

Let us begin with putting the CCWT into the context of quantum mechanics.
Based on the idea of quantum entanglement initiated by
Einstein-Podolsky-Rosen (EPR) \cite{13},\ Fan and Klauder constructed the
entangled state representation in two-mode Fock space \cite{8}, $\left\vert
\eta \right\rangle $ in (\ref{1.4}) is the common eigenvector of two
particles' relative position $X_{1}-X_{2}$ and their momentum $P_{1}+P_{2},$
\begin{equation}
\left( X_{1}-X_{2}\right) \left\vert \eta \right\rangle =\sqrt{2}\eta
_{1}\left\vert \eta \right\rangle ,\text{ }\left( P_{1}+P_{2}\right)
\left\vert \eta \right\rangle =\sqrt{2}\eta _{2}\left\vert \eta
\right\rangle ,  \label{1.9}
\end{equation}%
where $X_{j}=(a_{j}+a_{j}^{\dag })/\sqrt{2},$\ $P_{j}=(a_{j}-a_{j}^{\dag })/(%
\sqrt{2}i),$ ($j=1,2)$. $\left\vert \eta \right\rangle $ is complete $\int
\frac{d^{2}\eta }{\pi }\left\vert \eta \right\rangle \left\langle \eta
\right\vert =1$ ($d^{2}\eta \equiv d\eta _{1}d\eta _{2},$\ $\eta =\eta
_{1}+i\eta _{2}),$ and orthonormal $\left\langle \eta \right\vert \left.
\eta ^{\prime }\right\rangle =\pi \delta \left( \eta -\eta ^{\prime }\right)
\delta \left( \eta ^{\ast }-\eta ^{\prime \ast }\right) \equiv \pi \delta
^{(2)}\left( \eta -\eta ^{\prime }\right) $.

Using $\left\langle \eta \right\vert $ and $\psi \left( \eta \right)
=\left\langle \eta \right\vert \left. \psi \right\rangle $\ we can recast
the CCWT in (\ref{1.5}) as%
\begin{equation}
W_{\psi }g\left( \mu ,\kappa \right) =\left\langle \psi \right\vert
U_{2}\left( \mu ,\kappa \right) \left\vert g\right\rangle ,  \label{1.12}
\end{equation}%
and $U_{2}\left( \mu ,\kappa \right) $ is a two-mode squeezing-translating
operator, which has its natural expression in EPR entangled state
representation,
\begin{equation}
U_{2}\left( \mu ,\kappa \right) \equiv \frac{1}{\mu }\int \frac{d^{2}\eta }{%
\pi }\left\vert \frac{\eta -\kappa }{\mu }\right\rangle \left\langle \eta
\right\vert ,  \label{1.13}
\end{equation}%
when $\kappa =0,$ $U_{2}\left( \mu ,0\right) =S_{2}.$

\section{Parseval Theorem in the CCWT}

Now let us prove the Parseval theorem for CCWT,%
\begin{equation}
\int_{0}^{\infty }\frac{d\mu }{\mu ^{3}}\int \frac{d^{2}\kappa }{\pi }%
W_{\psi }g_{1}\left( \mu ,\kappa \right) W_{\psi }^{\ast }g_{2}\left( \mu
,\kappa \right) =C_{\psi }^{\prime }\int \frac{d^{2}\eta }{\pi }g_{2}^{\ast
}\left( \eta \right) g_{1}\left( \eta \right) ,  \label{16}
\end{equation}%
where $\kappa =\kappa _{1}+i\kappa _{2},$%
\begin{equation}
C_{\psi }^{\prime }=4\int_{0}^{\infty }\frac{d\left\vert \xi \right\vert }{%
\left\vert \xi \right\vert }\left\vert \psi \left( \xi \right) \right\vert
^{2}.  \label{17}
\end{equation}%
$\psi \left( \xi \right) $ is the Fourier transform of $\psi \left( \eta
\right) $, a mother wavelet. According to (\ref{1.13}) and (\ref{1.12}) the
quantum mechanical version of Parseval theorem should be%
\begin{equation}
\int_{0}^{\infty }\frac{d\mu }{\mu ^{3}}\int \frac{d^{2}\kappa }{\pi }%
\left\langle \psi \right\vert U_{2}\left( \mu ,\kappa \right) \left\vert
g_{1}\right\rangle \left\langle g_{2}\right\vert U_{2}^{\dagger }\left( \mu
,\kappa \right) \left\vert \psi \right\rangle =C_{\psi }^{\prime
}\left\langle g_{2}\right. \left\vert g_{1}\right\rangle ,  \label{18}
\end{equation}%
where $\psi \left( \eta \right) =\left\langle \eta \right\vert \left. \psi
\right\rangle $, so $\psi \left( \xi \right) =$ $\left\langle \xi
\right\vert \left. \psi \right\rangle ,$ $\left\vert \xi \right\rangle $ is
the conjugate state to $\left\vert \eta \right\rangle ,$%
\begin{eqnarray}
\left\vert \xi \right\rangle &=&\exp \left\{ -\frac{1}{2}\left\vert \xi
\right\vert ^{2}+\xi a_{1}^{\dagger }+\xi ^{\ast }a_{2}^{\dagger
}-a_{1}^{\dagger }a_{2}^{\dagger }\right\} \left\vert 00\right\rangle  \notag
\\
&=&\left( -1\right) ^{a_{2}^{\dagger }a_{2}}\left\vert \eta \right\rangle
_{\eta =\xi },\text{ \ \ \ }\xi =\xi _{1}+i\xi _{2},  \label{1.19}
\end{eqnarray}%
which is the common eigenstate of center-of-mass coordinate and the relative
momentum operators, i.e.,
\begin{equation}
\left( X_{1}+X_{2}\right) \left\vert \xi \right\rangle =\sqrt{2}\xi
_{1}\left\vert \xi \right\rangle ,\text{ }\left( P_{1}-P_{2}\right)
\left\vert \xi \right\rangle =\sqrt{2}\xi _{2}\left\vert \xi \right\rangle ,
\label{1.20}
\end{equation}%
and is complete%
\begin{equation}
\int \frac{d^{2}\xi }{\pi }\left\vert \xi \right\rangle \left\langle \xi
\right\vert =1.  \label{m}
\end{equation}%
The overlap between $\left\langle \xi \right\vert $ and $\left\vert \eta
\right\rangle $ is \cite{14}
\begin{equation}
\left\langle \xi \right. \left\vert \eta \right\rangle =\frac{1}{2}\exp [%
\frac{1}{2}\left( \xi ^{\ast }\eta -\xi \eta ^{\ast }\right) ]=\frac{1}{2}%
\exp \left[ \allowbreak i\left( \xi _{1}\eta _{2}-\xi _{2}\eta _{1}\right) %
\right] .  \label{1.22}
\end{equation}%
so%
\begin{eqnarray}
\psi \left( \xi \right) &=&\left\langle \xi \right\vert \left. \psi
\right\rangle =\int \frac{d^{2}\eta }{\pi }\left\langle \xi \right\vert
\left. \eta \right\rangle \left\langle \eta \right\vert \left. \psi
\right\rangle  \notag \\
&=&\int \frac{d^{2}\eta }{2\pi }\exp \left[ \left( \xi ^{\ast }\eta -\xi
\eta ^{\ast }\right) /2\right] \psi \left( \eta \right) .  \label{1.21}
\end{eqnarray}%
Eq.(\ref{16}) indicates that once the state vector $\left\langle \psi
\right\vert $ corresponding to mother wavelet is known, for any two states $%
\left\vert g_{1}\right\rangle $ and $\left\vert g_{2}\right\rangle $, their
overlap up to the factor $C_{\psi }$ (determined by (\ref{17})) is just
their corresponding overlap of CCWTs in the ($\mu ,\kappa $) parametric
space.

\textbf{Proof of Eq.(\ref{16}) or (\ref{18})}

We start with calculating $U_{2}^{\dagger }\left( \mu ,\kappa \right)
\left\vert \xi \right\rangle .$ Using (\ref{1.13}) and (\ref{1.22}), we have%
\begin{eqnarray}
U_{2}^{\dagger }\left( \mu ,\kappa \right) \left\vert \xi \right\rangle &=&%
\frac{1}{\mu }\int \frac{d^{2}\eta }{\pi }\left\vert \eta \right\rangle
\left\langle \frac{\eta -\kappa }{\mu }\right\vert \left. \xi \right\rangle
\notag \\
&=&\frac{1}{\mu }\int \frac{d^{2}\eta }{2\pi }\left\vert \eta \right\rangle
e^{\frac{i}{\mu }\left( \xi _{2}\eta _{1}-\xi _{1}\eta _{2}+\xi _{1}\kappa
_{2}-\xi _{2}\kappa _{1}\right) }  \notag \\
&=&\frac{1}{\mu }\left\vert \frac{\xi }{\mu }\right\rangle e^{\frac{i}{\mu }%
\left( \xi _{1}\kappa _{2}-\xi _{2}\kappa _{1}\right) },  \label{19}
\end{eqnarray}%
it follows%
\begin{eqnarray}
&&\int \frac{d^{2}\kappa }{\pi }U_{2}^{\dagger }\left( \mu ,\kappa \right)
\left\vert \xi ^{\prime }\right\rangle \left\langle \xi \right\vert
U_{2}\left( \mu ,\kappa \right)  \notag \\
&=&\frac{1}{\mu ^{2}}\int \frac{d^{2}\kappa }{\pi }e^{\frac{i}{\mu }\left[
\left( \xi _{1}^{\prime }-\xi _{1}\right) \kappa _{2}+\left( \xi _{2}-\xi
_{2}^{\prime }\right) \kappa _{1}\right] }\left\vert \frac{\xi ^{\prime }}{%
\mu }\right\rangle \left\langle \frac{\xi }{\mu }\right\vert  \notag \\
&=&4\pi \left\vert \frac{\xi ^{\prime }}{\mu }\right\rangle \left\langle
\frac{\xi }{\mu }\right\vert \delta \left( \xi _{1}^{\prime }-\xi
_{1}\right) \delta \left( \xi _{2}-\xi _{2}^{\prime }\right) .  \label{20}
\end{eqnarray}%
Using (\ref{m}) and (\ref{20}) the left-hand side (LHS) of (\ref{18}) can be
reformed as%
\begin{eqnarray}
&&\text{LHS of Eq.(\ref{18})}  \notag \\
&=&\int_{0}^{\infty }\frac{d\mu }{\mu ^{3}}\int \frac{d^{2}\kappa d^{2}\xi
d^{2}\xi ^{\prime }}{\pi ^{3}}\left\langle \psi \right\vert \left. \xi
\right\rangle  \notag \\
&&\times \left\langle \xi \right\vert U_{2}\left( \mu ,\kappa \right)
\left\vert g_{1}\right\rangle \left\langle g_{2}\right\vert U_{2}^{\dagger
}\left( \mu ,\kappa \right) \left\vert \xi ^{\prime }\right\rangle
\left\langle \xi ^{\prime }\right\vert \left. \psi \right\rangle  \notag \\
&=&4\int_{0}^{\infty }\frac{d\mu }{\mu ^{3}}\int \frac{d^{2}\xi d^{2}\xi
^{\prime }}{\pi }\left\langle g_{2}\right. \left\vert \frac{\xi ^{\prime }}{%
\mu }\right\rangle \left\langle \frac{\xi }{\mu }\right. \left\vert
g_{1}\right\rangle  \notag \\
&&\times \psi ^{\ast }\left( \xi \right) \psi \left( \xi ^{\prime }\right)
\delta \left( \xi _{1}^{\prime }-\xi _{1}\right) \delta \left( \xi _{2}-\xi
_{2}^{\prime }\right)  \notag \\
&=&4\int_{0}^{\infty }\frac{d\mu }{\mu ^{3}}\int \frac{d^{2}\xi }{\pi }%
\left\vert \psi \left( \xi \right) \right\vert ^{2}\left\langle g_{2}\right.
\left\vert \frac{\xi }{\mu }\right\rangle \left\langle \frac{\xi }{\mu }%
\right. \left\vert g_{1}\right\rangle  \notag \\
&=&\int \frac{d^{2}\xi }{\pi }\left\{ 4\int_{0}^{\infty }\frac{d\mu }{\mu }%
\left\vert \psi \left( \mu \xi \right) \right\vert ^{2}\right\} \left\langle
g_{2}\right. \left\vert \xi \right\rangle \left\langle \xi \right.
\left\vert g_{1}\right\rangle ,  \label{22}
\end{eqnarray}%
where the integration value in $\{..\}$ is actally $\xi -$independent.
Noting that the mother wavelet $\psi \left( \eta \right) $ in Eq.(\ref{23})
is just the function of $\left\vert \eta \right\vert ,$ so $\psi \left( \xi
\right) $ is also the function of $\left\vert \xi \right\vert .$ In fact,
using Eqs.(\ref{23}),(\ref{24}) and (\ref{1.21}), we have
\begin{equation}
\psi \left( \xi \right) =e^{-1/2\left\vert \xi \right\vert
^{2}}\sum_{n=0}^{\infty }K_{n,n}H_{n,n}\left( \left\vert \xi \right\vert
,\left\vert \xi \right\vert \right) ,  \label{25}
\end{equation}%
where we have used the integral formula%
\begin{equation}
\int \frac{d^{2}z}{\pi }e^{\zeta \left\vert z\right\vert ^{2}+\xi z+\eta
z^{\ast }}=-\frac{1}{\zeta }e^{-\frac{\xi \eta }{\zeta }},\text{Re}\left(
\zeta \right) <0.  \label{26}
\end{equation}%
So we can rewrite (\ref{22}) as
\begin{equation}
\text{LHS of (\ref{18})}=C_{\psi }^{\prime }\int \frac{d^{2}\xi }{\pi }%
\left\langle g_{2}\right. \left\vert \xi \right\rangle \left\langle \xi
\right. \left\vert g_{1}\right\rangle =C_{\psi }^{\prime }\left\langle
g_{2}\right. \left\vert g_{1}\right\rangle ,  \label{27}
\end{equation}%
where%
\begin{equation}
C_{\psi }^{\prime }=4\int_{0}^{\infty }\frac{d\mu }{\mu }\left\vert \psi
\left( \mu \xi \right) \right\vert ^{2}=4\int_{0}^{\infty }\frac{d\left\vert
\xi \right\vert }{\left\vert \xi \right\vert }\left\vert \psi \left( \xi
\right) \right\vert ^{2}.  \label{28}
\end{equation}%
Then we have completed the proof of the Parseval theorem for CCWT in (\ref%
{18}). Here, we should emphasize that (\ref{18}) is not only different from
the product of two 1D WTs, but also different from the usual wavelet
transform in 2D.

When $\left\vert g_{2}\right\rangle =\left\vert \eta \right\rangle ,$ by
using (\ref{1.13}) we see
\begin{equation}
\left\langle \eta \right\vert U_{2}^{\dagger }\left( \mu ,\kappa \right)
\left\vert \psi \right\rangle =\frac{1}{\mu }\psi \left( \frac{\eta -\kappa
}{\mu }\right) ,  \label{29}
\end{equation}%
then substituting it into (\ref{18}) yields%
\begin{equation}
g_{1}\left( \eta \right) =\frac{1}{C_{\psi }^{\prime }}\int_{0}^{\infty }%
\frac{d\mu }{\mu ^{3}}\int \frac{d^{2}\kappa }{\pi \mu }W_{\psi }g_{1}\left(
\mu ,\kappa \right) \psi \left( \frac{\eta -\kappa }{\mu }\right) ,
\label{30}
\end{equation}%
which is just the inverse transform of the CCWT.

Especially, when $\left\vert g_{1}\right\rangle =$ $\left\vert
g_{2}\right\rangle ,$ (\ref{18}) reduces to%
\begin{eqnarray}
\int_{0}^{\infty }\frac{d\mu }{\mu ^{3}}\int \frac{d^{2}\kappa }{\pi }%
\left\vert W_{\psi }g_{1}\left( \mu ,\kappa \right) \right\vert ^{2}
&=&C_{\psi }^{\prime }\int \frac{d^{2}\eta }{\pi }\left\vert g_{1}\left(
\eta \right) \right\vert ^{2},  \notag \\
\text{or }\int_{0}^{\infty }\frac{d\mu }{\mu ^{3}}\int \frac{d^{2}\kappa }{%
\pi }\left\vert \left\langle \psi \right\vert U_{2}\left( \mu ,\kappa
\right) \left\vert g_{1}\right\rangle \right\vert ^{2} &=&C_{\psi }^{\prime
}\left\langle g_{1}\right. \left\vert g_{1}\right\rangle ,  \label{31}
\end{eqnarray}%
which is named isometry of energy.

\section{New orthogonal property of mother wavelet in parameter space}

On the other hand, when $\left\vert g_{1}\right\rangle =\left\vert \eta
\right\rangle ,$ $\left\vert g_{2}\right\rangle =\left\vert \eta ^{\prime
}\right\rangle $, (\ref{18}) becomes%
\begin{equation}
\frac{1}{C_{\psi }^{\prime }}\int_{0}^{\infty }\frac{d\mu }{\mu ^{5}}\int
\frac{d^{2}\kappa }{\pi }\psi \left( \frac{\eta ^{\prime }-\kappa }{\mu }%
\right) \psi ^{\ast }\left( \frac{\eta -\kappa }{\mu }\right) =\pi \delta
^{(2)}\left( \eta -\eta ^{\prime }\right) ,  \label{32}
\end{equation}%
which is a new orthogonal property of mother wavelet in parameter space
spanned by $\left( \mu ,\kappa \right) $. In a similar way, we take $%
\left\vert g_{1}\right\rangle =\left\vert g_{2}\right\rangle =\left\vert
m,n\right\rangle ,$ a two-mode number state, since $\left\langle m,n\right.
\left\vert m,n\right\rangle =1,$ then we have
\begin{equation}
\int_{0}^{\infty }\frac{d\mu }{\mu ^{3}}\int \frac{d^{2}\kappa }{\pi }%
\left\vert \left\langle \psi \right\vert U_{2}\left( \mu ,\kappa \right)
\left\vert m,n\right\rangle \right\vert ^{2}=C_{\psi }^{\prime },
\label{1.37}
\end{equation}%
or take $\left\vert g_{1}\right\rangle =\left\vert g_{2}\right\rangle
=\left\vert z_{1},z_{2}\right\rangle ,$ $\left\vert z\right\rangle =\exp
\left( -\left\vert z\right\vert ^{2}/2+za^{\dagger }\right) \left\vert
0\right\rangle $ is the coherent state, then
\begin{equation}
\int_{0}^{\infty }\frac{d\mu }{\mu ^{3}}\int \frac{d^{2}\kappa }{\pi }%
\left\vert \left\langle \psi \right\vert U_{2}\left( \mu ,\kappa \right)
\left\vert z_{1},z_{2}\right\rangle \right\vert ^{2}=C_{\psi }^{\prime }.
\label{1.38}
\end{equation}%
This indicates that $C_{\psi }^{\prime }$ is $\left\vert g_{1}\right\rangle $%
-independent, which coincides with the expression in (\ref{17}).

Next we examine a special example. When the mother wavelet $\psi \left( \eta
\right) $ is taken as the following form%
\begin{equation}
\psi _{M}\left( \eta \right) =\left\langle \eta \right. \left\vert \psi
\right\rangle =e^{-1/2\left\vert \eta \right\vert ^{2}}(1-\frac{1}{2}%
\left\vert \eta \right\vert ^{2}),  \label{33}
\end{equation}%
which is different from $e^{-(x^{2}+y^{2})/2}(1-x^{2})\left( 1-y^{2}\right)
, $ the direct-product of two 1D Mexican hat wavelets (we name entangled
mexican hat wavelets (EMHWs)), using (\ref{1.21}) we have%
\begin{equation}
\psi \left( \xi \right) =\frac{1}{2}\left\vert \xi \right\vert ^{2}e^{-\frac{%
1}{2}\left\vert \xi \right\vert ^{2}},  \label{34}
\end{equation}%
which leads to
\begin{equation}
C_{\psi }^{\prime }=\int_{0}^{\infty }\left\vert \xi \right\vert
^{3}e^{-\left\vert \xi \right\vert ^{2}}d\left\vert \xi \right\vert =\frac{1%
}{2}.  \label{35}
\end{equation}%
Thus for the EMHWs (\ref{33}), we see%
\begin{equation}
2\int_{0}^{\infty }\frac{d\mu }{\mu ^{5}}\int \frac{d^{2}\kappa }{\pi }\psi
_{M}\left( \frac{\eta ^{\prime }-\kappa }{\mu }\right) \psi _{M}^{\ast
}\left( \frac{\eta -\kappa }{\mu }\right) =\pi \delta ^{(2)}\left( \eta
-\eta ^{\prime }\right) .  \label{36}
\end{equation}%
Eq. (\ref{36}) can be checked as follows. Using (\ref{33}) and the integral
formula%
\begin{eqnarray}
&&\int_{0}^{\infty }u\left( 1-\frac{ux^{2}}{2}\right) \left( 1-\frac{uy^{2}}{%
2}\right) e^{-u\frac{x^{2}+y^{2}}{2}}du  \notag \\
&=&-\frac{4(x^{4}-4x^{2}y^{2}+y^{4})}{(x^{2}+y^{2})^{4}},\text{ Re}\left(
x^{2}+y^{2}\right) >0,  \label{38}
\end{eqnarray}%
we can put the left-hand side (LHS) of (\ref{36}) into
\begin{eqnarray}
&&\text{LHS of (\ref{36})}  \notag \\
&=&-\int_{0}^{\infty }\frac{d\frac{1}{\mu ^{2}}}{\mu ^{2}}\int \frac{%
d^{2}\kappa }{\pi }e^{-\frac{x^{2}+y^{2}}{2\mu ^{2}}}\left( 1-\frac{x^{2}}{%
2\mu ^{2}}\right) \left( 1-\frac{y^{2}}{2\mu ^{2}}\right)  \notag \\
&=&\int_{0}^{\infty }udu\int \frac{d^{2}\kappa }{\pi }\left( 1-\frac{ux^{2}}{%
2}\right) \left( 1-\frac{uy^{2}}{2}\right) e^{-u\frac{x^{2}+y^{2}}{2}}
\notag \\
&=&-\int \frac{d^{2}\kappa }{\pi }\frac{4(x^{4}-4x^{2}y^{2}+y^{4})}{%
(x^{2}+y^{2})^{4}},  \label{37}
\end{eqnarray}%
where $x^{2}=\left\vert \eta ^{\prime }-\kappa \right\vert ^{2},$ $%
y^{2}=\left\vert \eta -\kappa \right\vert ^{2}.$

When $\eta ^{\prime }=\eta ,$ $x^{2}=y^{2},$

\begin{equation}
\text{LHS of (\ref{36})}=\int \allowbreak \frac{d^{2}\kappa }{2\pi
\left\vert \kappa -\eta \right\vert ^{4}}=\int_{0}^{\infty }\allowbreak
\int_{0}^{2\pi }\frac{drd\theta }{2\pi r^{3}}\rightarrow \infty .  \label{39}
\end{equation}%
On the other hand, when $\eta \neq \eta ^{\prime }$ and noticing that
\begin{eqnarray}
x^{2} &=&\left( \eta _{1}^{\prime }-\kappa _{1}\right) ^{2}+\left( \eta
_{2}^{\prime }-\kappa _{2}\right) ^{2},  \notag \\
y^{2} &=&\left( \eta _{1}-\kappa _{1}\right) ^{2}+\left( \eta _{2}-\kappa
_{2}\right) ^{2},  \label{40}
\end{eqnarray}%
which leads to
\begin{equation}
\mathtt{d}x^{2}\mathtt{d}y^{2}=4\left\vert J\right\vert \mathtt{d}\kappa _{1}%
\mathtt{d}\kappa _{2}\text{,}  \label{41}
\end{equation}%
where $J\left( x,y\right) =\left\vert
\begin{array}{cc}
\kappa _{1}-\eta _{1}^{\prime } & \kappa _{2}-\eta _{2}^{\prime } \\
\kappa _{1}-\eta _{1} & \kappa _{2}-\eta _{2}%
\end{array}%
\right\vert $. As a result of (\ref{40}), (\ref{37}) reduces to
\begin{equation}
\text{LHS of (\ref{36})}=-4\int_{-\infty }^{\infty }\frac{dxdy}{\pi }\frac{%
xy(x^{4}-4x^{2}y^{2}+y^{4})}{\left\vert J\right\vert (x^{2}+y^{2})^{4}}=0%
\text{,}  \label{42}
\end{equation}%
where we have noticed that $J\left( x,y\right) $ is the funtion of $\left(
x^{2},y^{2}\right) .$ Thus we have%
\begin{equation}
\text{LHS of (\ref{36})}=\left\{
\begin{array}{cc}
\infty , & \eta =\eta ^{\prime }, \\
0, & \eta \neq \eta ^{\prime }.%
\end{array}%
\right. =\text{RHS of (\ref{36}).}  \label{43}
\end{equation}

In sum, we have proposed the Parseval theorem corresponding to the CCWT in
the context of quantum mechanics. Our calculations are simplified greatly by
using the quantum state representations of two-mode squeezing operators.
Finally, we should emphasize that since the CCWT corresponds to two-mode
squeezing transform which differs from two single-mode squeezing operators'
direct product, the Parseval theorem of CCWT defined in this paper differs
from that of the direct product of two 1-dimensional wavelet transforms.

\textbf{Acknowledgements}

This work was supported by the National Natural Science Foundation of China
(Grant Nos. 10775097), and the Research Foundation of the Education
Department of Jiangxi Province.


\begin{thebibliography}{99}
\bibitem{1} S. Jaffard, Y. Meyer, and R. D. Ryan, Wavelets, Tools for
Science \& Technology (Society for Industrial and Applied Mathematics, 2001)

\bibitem{2} I. Daubechies, Ten Lectures on Wavelets, CBMS-NSF Series in
Applied Mathematics (SIAM, 1992).

\bibitem{3} M. A. Pinsky, Introduction to Fourier Analysis and Wavelets
(Book/Cole, 2002).

\bibitem{4} J. W. Goodman, Introduction to Fourier Optics, (McGraw-Hill, New
York, 1968).

\bibitem{5} L.-Y. Hu and H.-Y. Fan, Wavelet transform in the context of
quantum mechanics and new orthogonal property of mother wavelets in
parameter space, J. Mod. Opt. \textbf{55}, 1835-1844 (2008).

\bibitem{6} H.-Y. Fan and H.-L. Lu, General formula for finding mother
wavelets by virtue of Dirac's representation theory and the coherent state,
Opt. Lett. \textbf{31}, 407-409 (2006),

\bibitem{7} H.-Y. Fan and J.-F Lu ,\ Quantum Mechanics Version of Wavelet
Transform studied by virtue of IWOP technique, Commun. Theor. Phys. \textbf{%
41}, 681-684 (2004).

\bibitem{8} H.-Y. Fan and J. R. Klauder, Eigenvectors of two particles'
relative position and total momentum Phys. Rev. A 49, 704-707 (1994).

\bibitem{9} H.-Y. Fan and H.-L. Lu, Mother wavelet for complex wavelet
transform derived by EPR entangled state representation, Opt. Lett. 32,
554-556 (2007).

\bibitem{10} A. Wunsche, General Hermite and Laguerre two-dimensional
polynomials, J. Phys. A-Math. and Gen. \textbf{33}, 1603-1629 (2000).

\bibitem{11} R. Loudon, and P.L. Knight, Squeezed light, J. Mod. Opt. 34,
709-759 (1987).

\bibitem{12} M. O. Scully and M. S. Zubairy, Quantum Optics, (Cambridge
University, 1997).

\bibitem{13} A. Einstein, B. Podolsky and N. Rosen, Can quantum-mechanical
description of physical reality be considered complete? Phys. Rev. 47,
777-780 (1935).

\bibitem{14} H.-Y. Fan and H.-L. Lu, Classical optical transforms studied in
the context of quantum optics via the route of developing Dirac's symbolic
method, Int. J. Mod. Phys. B, 19, 799-856 (2005).
\end{thebibliography}
\end{document}